\documentclass[prl, aps, twocolumn, superscriptaddress, nofootinbib, tightenlines, nobibnotes, showpacs, 10pt]{revtex4-1}

\usepackage{natbib}
\usepackage{slashed}
\usepackage{graphicx}
\usepackage{subfigure}
\usepackage[usenames, dvipsnames]{color}
\usepackage{graphics}
\usepackage{hyperref}
\usepackage{bm}
\usepackage{amsmath}
\usepackage{color}
\usepackage{amsfonts}
\hypersetup{backref,
colorlinks=true,
linkcolor=blue,
linktoc=page,
citecolor=blue,
urlcolor=blue}

\everymath{\displaystyle}

\begin{document}

\title{Towards the detection of ultra-low energetic neutrinos with plasma metamaterials}

\author{Carlo Alfisi}
\email{carlo1.alfisi@mail.polimi.it}
\affiliation{Politecnico di Milano, Milan, Italy}
\affiliation{Instituto Superior T\'ecnico, Lisboa, Portugal}

\author{Hugo Ter\c{c}as}
\email{hugo.tercas@tecnico.ulisboa.pt}
\affiliation{GoLP/Instituto de Plasmas e Fus\~ao Nuclear, Lisbon, Portugal}
\affiliation{Instituto Superior T\'ecnico, Lisbon, Portugal}

\begin{abstract}

Experiments as IceCube or Super-Kamiokande have been successful in detecting highly energetic neutrinos in the. Neutrinos in the ultra-low energy range ($\mathcal{E}<1.0~\rm{eV}$) have been theoretically predicted but their observation remain elusive, and no concrete experimental scheme has been proposed for that job. Here, we propose a novel scheme based on graphene plasmonic metamaterials to designed to detect ultra-low energetic neutrinos. We claim that slow neutrino fluxes, interacting with solid-state plasmas, can generate an instability due to the weak neutrino-plasmon interaction, which is reminiscent of the beam-plasma instability taking place in astrophysics and laboratory plasmas. We make use of the semi-classical limit of the weak interaction to describe the coupling between the neutrinos and electrons in graphene. To render the scheme practical, we investigate the neutrino-plasma instability produced in a graphene metamaterial, composed by a periodic stacking of graphene layers. Our findings reveal that the controlled excitation of plasma waves in such graphene metamaterial allows for the detection of neutrinos in the energy range $\sim 1.0~\rm{\mu eV}-100~\rm{meV}$, and fluxes in the range $10^{4}-10^{10} \rm{cm^{-2} s^{-1}}$.
\end{abstract}
\maketitle


\textit{Introduction.} The scientific community is facing problems as the extension of the Standard Model \cite{bilenky:nsmb}, the unknown Dark Matter \cite{feng:dmcppmd}, Cosmogenesis \cite{joyce:bcsm}, Supernova explosion \cite{fuller:hsnse} and many others. Several theories have been proposed but only experimental data can confirm their validity, one promising benchmark is studying the neutrino properties \cite{giunti:fnpa}.
In 2020 it was proposed the neutrino spectra that reaches Earth, which is obtained summing over the three flavor and integrating over all directions \cite{vitagliano:gunsae}. The result shows that neutrinos can have energy from $\rm{\mu eV}$ up to $\rm{PeV}$. The high energetic neutrinos are detected by several experiments \cite{IMB:exp,SuperKamio:exp,SNO:exp,halzen:IC}; on the contrary, the neutrinos at \textit{ultra-low energy}, i.e. with energies in the range $\rm{\mu eV}-\rm{eV}$, remain elusive up to today. The aim of this work is to propose a way to detect slow neutrino fluxes. It is well established that neutrinos release a large fraction to their energy to the plasma during the collapse of stars, such that neutrino-plasma instabilities may eventually be a crucial ingredient to supernova explosions \cite{silva:cnpi,bingham:cibnadp,serbeto:ndwiepp}. Inspired by this mechanism, we seek for a scheme to enhance the interaction of neutrino fluxes with a solid-state plasma. The interaction may lead to an instability which excites and amplifies the plasmons, and the resulting plasmon may then be used to infer about the neutrino properties. In this Letter, we propose to use a metamaterial composed by a periodic stacking of graphene field-effect transistors (FETs) as the buffer medium (detector). The reason for this choice stems in the long lifetime of the plasmons in graphene, as consequence of the high mobility in electrons close to the Dirac points \cite{yao:bgtpgh,morgado:nldbg,grigorenko:gp}. 

\begin{figure}[t!]
	\centering
	\includegraphics[width=0.4\textwidth]{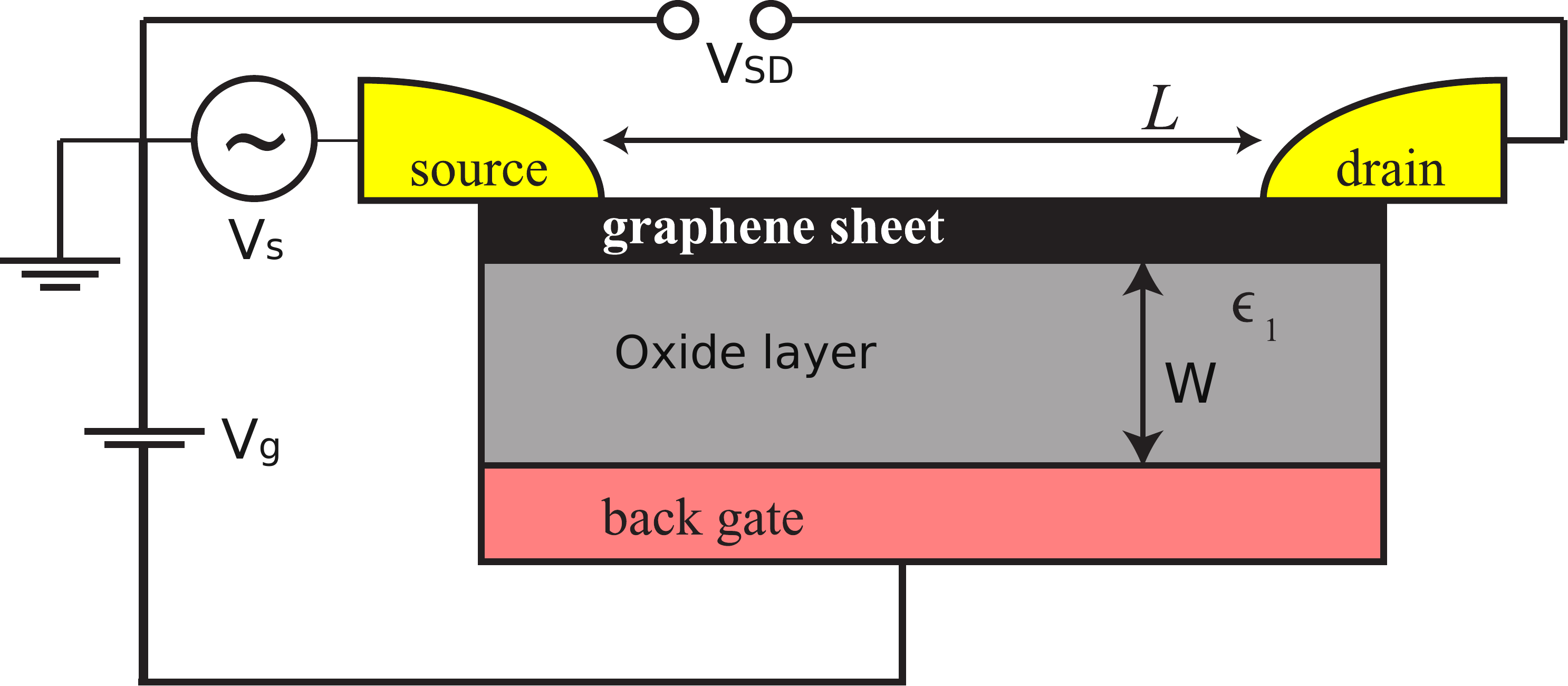}
	\caption{The field-effect transistor is composed by a graphene channel (length $L$) at which edges there are source and drain contacts, an oxide layer (thickness $W$) separate the channel to gate contact.}
	\label{fig:FET}
\end{figure}

\textit{Plasmons in graphene field-effect transistors.} The solid-state plasma device that we use as element consists in a graphene field-effect transistor (FET), as schematically represented in Fig. \ref{fig:FET}. Near the Dirac points, electron have a relativistic-like dispersion $\epsilon=\hbar v_F k$, where $v_F$ is the slope of the Dirac cone. In such a scenario, and in the limit $k_BT\ll E_F$, with $E_F$ denoting the Fermi level $-$ which is ultimately controlled by the gate voltage $V_g$ $-$ the electron effective mass is given by its Drude value, $M\simeq \hbar\sqrt{\pi N_e}/v_F$, with $N_e$ denoting the equilibrium density. Here, the electrons are considered to be initially at rest, $\mathbf{V}_e=0$.
The electron density oscillations $-$plasmons$-$ can be generated by imposing a voltage modulation at the source contact \cite{bandurin:rtdugp}, or by the \textit{Dyakonov-Shur instability} \cite{dyakonov:pwe}, or even by coupling an electromagnetic radiation \cite{mashinsky:gptdhr,dai:taffmkrc}. The generated plasma wave propagates along the channel following the \textit{dispersion relation}. Such relation can be tuned by the boundary conditions, geometrical sizes and electric parameters. The resulting plasmons have frequency in the order of $\rm{THz}$ \cite{ryzhii:first,ryzhii:second,ryzhii:third}; therefore, the corresponding energy lies in the range $\mathcal{E}\equiv \hbar \omega \gtrsim 1~\rm{meV}$. The later is exactly the energy range of the neutrinos we are interested in. The plasma wave can be detected, the optimal technique should ensure the evaluation of the wave parameters: $k$ and $\omega$. Asymmetrical boundary conditions, such as open (closed) circuit at the source (drain), result in a dc potential across the channel. In the case the plasmon propagates across the channel without significant attenuation, the source-to-drain voltage is peaked when the plasmon has a frequency about the odd harmonics of the channel ($\Omega_j=(2j + 1) \pi S / 2L$). We can write it by the \textit{responsivity} as \cite{dyakonov:pwe}
\begin{equation}
	\mathcal{R}(\omega) = \left(\frac{S \tau}{L}\right)^2 \frac{(V_0 / V_{\rm{g}})^2}{4 (\omega-\Omega_j)^2 \tau^2 +1} ,\label{eq:DSresponce}
\end{equation}
where $V_0$ is the amplitude of the voltage modulation at the source, the point of the device at which the plasmon is generated. The relaxation time $\tau$ is related to the plasmon decay. The decay could occurs due to the scattering with the electrons and, if a modification of the graphene device is implemented, this induces a thermal current with spatial modulation proportional to the plasmon wavelength \cite{lundeberg:tdipgp}.
Moreover, an electromagnetic radiation is emitted due to the variations on the plasmon density and velocity, and collecting such a radiation allows to perform a spectral analysis of the plasmon. Thus, we can measure the frequency of the plasma oscillation \cite{tercas:appsmp}. The signal-to-noise ratio is defined as
\begin{equation}
	\rm{SN}_{\rm{dB}}=10 \log_{10}\left(\frac{A_s^2}{A_n^2}\right), \label{eq:SN}
\end{equation}
where $A_s$ ($A_n$) is the amplitude of the signal (noise). Regardless the specific technique in use, we need a signal-to-noise ratio larger than 1 to be able to measure the plasma waves. This condition will set the lower bounds for the neutrino flux we can detect.

\textit{Weak interaction.} The neutrino ($\nu$) is a chargeless lepton, and a precise measurement of its mass $m$ remains an open question, although experimental data constrains its value to the range $m<1~\rm{eV}$ \cite{loureiro:ubnm}. Neutrinos can interact with the electrons by means of the \textit{weak force} \cite{bachall:na}, which couples leptons through the Fermi constant $G_{\rm{F}} \simeq 1.4 \times 10^{-36}\, \rm{J\, m^3}$. The single-particle interaction is described by the quantum field Lagrangian as \cite{silva:cnpi}
\begin{eqnarray}
	& \mathcal{L}^{\rm{int}} = -\frac{G_{\rm{F}}}{\sqrt{2}} \bar{\psi}_e \gamma^\mu \big[(1- \gamma_5) + (C_{\rm{V}} - C_{\rm{A}} \gamma_5)\big]\psi_e \times \nonumber \\
	& \;\;\;\;\;\;\;\;\;\;\;\;\;\;\;\;\;\;\;\; \times \; \bar{\psi}_\nu \gamma_\mu (1- \gamma_5)\psi_\nu , \label{eq:lagrangianQ}
\end{eqnarray}
where $\psi_e$ ($\psi_\nu$)
is the quantum electron (neutrino) field, $\gamma_5 = i \gamma_0\gamma_1\gamma_2\gamma_3$ with $\gamma_\mu$ ($\mu$ = \{0, 1, 2, 3\}) being the Dirac matrices, and $C_{\rm{V}}$ and $C_{\rm{A}}$ respectively standing for the coupling constants for the vector and the axial currents. Eq. \eqref{eq:lagrangianQ} can be reduced to a semiclassical Lagrangian where the interaction takes place between a collection of neutrinos and electrons, described by macroscopic variables of density and a velocity fields. The semi-classical theory is obtained by transforming the quantum fields in quadrivectors as $\bar{\psi}\gamma_\mu\psi \rightarrow (n,n\,\mathbf{v}/c)$ and $\bar{\psi}\gamma_\mu(1-\gamma_5)\psi \rightarrow (n,n\,\mathbf{v}/c)$, which yields
\begin{equation}
	\mathcal{L}^{\rm{int}} = -\frac{G_{\rm{F}}}{\sqrt{2}}(C_{\rm{V}}+1) \left[n_e n_\nu -\frac{(n_e \mathbf{v}_e) \cdot (n_\nu \mathbf{v}_\nu)}{c^2} \right]. \label{eq:lagrangianSC}
\end{equation}
We can derive the weak neutrino-electron forces with the help of a semi-classical Hamiltonian, which can be obtained by the total Lagrangian, $\mathcal{L}=\mathcal{L}^0 + \mathcal{L}^{\rm{int}}$, where $\mathcal{L}^0$ is the free Lagrangian for an electron distribution and a relativistic neutrino flux. The conjugate momentum 
\begin{equation}
\label{eq:conjugateMomentum}
\mathbf{P}_i =\frac{\partial \mathcal{L}}{\partial{\dot{\mathbf{x}}}} =  \mathbf{p}_i + \tilde{G}_{\rm{F}} \frac{n_e n_\nu \mathbf{v}_i}{c^2},
\end{equation}
with $\mathbf{p}_i$ being the momentum of the species $i$ and $\tilde{G}_{\rm{F}} = G_{\rm{F}}(C_{\rm{V}}+1)/\sqrt{2}$, together with the Legendre transformation \cite{doughty:li}, yields
\begin{equation}
	H(P_e,P_\nu,x)= \sum \mathbf{P}_i \cdot \dot{\mathbf{x}}_i - \mathcal{L}.\label{eq:legendreTransf}
\end{equation}
The latter allows us to compute the force experienced by a the two species $i=\{e,\nu\}$ due to the weak interaction with the distribution of the species $j=\{e,\nu\}$ \cite{silva:cnpi},
\begin{equation}
	\mathbf{F}^W_i = -\tilde{G}_{\rm{F}} \left[ \mathbf{\nabla} n_j + \frac{\partial_t (n_j \mathbf{v}_j)}{c^2} + \frac{\mathbf{v}_i \wedge \mathbf{\nabla} \wedge (n_j \mathbf{v}_j)}{c^2} \right]. \label{eq:weakForce}
\end{equation}
In the following calculations, the last term in Eq. \eqref{eq:weakForce} will be neglected, as the typical spatial variations of the electron and neutrino densities are much smaller that the corresponding temporal ones. Notice that the electrons are electrical charged therefore, we will consider the electrostatic force $\mathbf{F}^{\rm EM}_e = e {\bm \nabla} \phi$ (where $\phi$ is electric potential, and we consider that no magnetic field is present) acting on the electrons. We aim to obtaining the joint modes of the graphene electrons and neutrino flux. By means of a kinetic theory, we write the following fluid model governing the neutrino-plasma interaction
\begin{eqnarray}
	& \partial_t n_e + \mathbf{\nabla} \cdot (n_e \mathbf{v}_e) = 0, \label{eq:continuityElectron} \\
	& \partial_t \mathbf{v}_e + \mathbf{v}_e \cdot \mathbf{\nabla} \mathbf{v}_e = \frac{\mathbf{F}^{\rm{EM}}+ \mathbf{F}^W_e}{M} , \label{eq:velocityContinuityElectron}\\
	& \partial_t n_\nu + \mathbf{\nabla} \cdot (n_\nu \mathbf{v}_\nu) = 0 , \label{eq:continuityNeutrino}\\
	&  \partial_t \mathbf{v}_\nu + \mathbf{v}_\nu \cdot \mathbf{\nabla} \mathbf{v}_\nu = \frac{ \mathbf{F}^W_\nu}{m} .\label{eq:velocityContinuityNeutrino}
\end{eqnarray}
The electron-neutrino modes are obtained by the joint dispersion relation which can be obtained by writing the matrix formulation of the problem and imposing the matrix determinant to vanish \cite{chen:ippcf}. To do so, we linearize Eqs. \eqref{eq:continuityElectron}$-$\eqref{eq:velocityContinuityNeutrino} as 
\begin{eqnarray}
	&& n_e(x,t) = N_e + n'_e(x,t), \\
	&& \mathbf{v}_e = \mathbf{v}'_e(x,t), \\
	&& n_\nu(x,t) = N_\nu + n'_\nu(x,t), \\
	&& \mathbf{v}_\nu(x,t) = \mathbf{V}_\nu + \mathbf{v}'_\nu(x,t),
\end{eqnarray}
where the perturbation terms $-$ labeled with the apostrophe $-$ must be smaller than the equilibrium quantities $\{N_e, \, N_\nu, \, \mathbf{V}_\nu \}$, which are assumed constant in space and time. We proceed by eliminating the velocity terms $\mathbf{v}'_e$ and $\mathbf{v}'_\nu$ obtaining two second-order differential equations for the electron and neutrino densities. We then move to the \textit{Fourier space} by the formal substitutions
\begin{equation}
	\partial_t \to -i \, \omega \;\; {\rm{and}} \;\; \mathbf{\nabla} \to +i \, k,
\end{equation}
where we implicitly stated that $\{n'_i, \phi \} \propto \exp[i(k x- \omega t)]$.
We are close to write the matrix form of the system but we still need to define the relation between electric potential and electron density; it depends on the value of the gate voltage. If the gate voltage is finite, there is a local relation, $\phi = -e n'_e / C$ with $C$ being the capacitance of the gate-oxide-channel structure; we call it \textit{gated electrons}. When the gate voltage is null, we can use the \textit{Poisson equation} thus $\phi = -e n'_e / 2 \epsilon_0 k$, namely \textit{ungated electrons}. In the following the apostrophe will be dropped to simplify the notation and $n_e$ ($n_\nu$) should be read as the amplitude of the perturbation of electron (neutrino) density in Fourier space. 
The different dimensionality of neutrino and electron densities must be considered; the electrons are confined in the bidimentional graphene channel which has thickness $a$, thus by \textit{dimensional reduction} we rewrite the proportionality constant of $F^W_\nu$ as
\begin{equation}
	\tilde{g}_{\rm{F}} \simeq \tilde{G}_{\rm{F}}/a.
\end{equation}
The gated case implies bare electron modes with a linear behavior, the corresponding matrix system for the interacting electron and neutrino distributions reads as
\begin{equation}
	\begin{bmatrix}
		\omega^2 - S^2 k^2  & \frac{N_e \tilde{G}_{\rm{F}} }{M c^2} \left( \omega^2- c^2 k^2 \right)\\[10 pt]
		\frac{N_\nu \tilde{g}_{\rm{F}} }{m c^2} \left( \omega^2- c^2 k^2 \right) & (\omega- V_\nu k )^2
	\end{bmatrix} 
	\cdot
	\begin{bmatrix}
		n_e  \\[10 pt]
		n_\nu
	\end{bmatrix} =0,
\end{equation}
where we introduced the plasmon speed $S=\sqrt{N_e e^2 /M C}$. In this case, the neutrino and plasmon modes do not exchange energy resonantly, i.e. they do not cross. Conversely, for ungated electrons, the bare plasma dispersion relation is $\omega\simeq \sqrt{g_e k}$, with $g_e=N_e e^2 /2 \epsilon_0 M$, and it crosses the linear neutrino dispersion, allowing for resonant mode conversion. The full electron-neutrino dispersion can be recast as
\begin{equation}
	\begin{bmatrix}
		\omega^2 - g_e k & \frac{N_e \tilde{G}_{\rm{F}} }{M c^2} \left( \omega^2 - c^2 k^2 \right)\\
		\frac{N_\nu \tilde{g}_{\rm{F}} }{m c^2} \left( \omega^2 - c^2 k^2 \right) & (\omega- V_\nu k )^2
	\end{bmatrix} 
	\cdot
	\begin{bmatrix}
		n_e  \\
		n_\nu
	\end{bmatrix} =0 , \label{eq:matrixUngated}
\end{equation}
The weak interaction bends the free modes avoiding the crossing resulting in complex modes. The imaginary part of the frequency ($\gamma$) implies an exponential growth of the plasmons amplitude i.e., an instability feature; similar behavior is found in the \textit{beam-plasma instability} \cite{gupta:nscbpi}.

\begin{figure}[t!]
\includegraphics[width=0.45\textwidth]{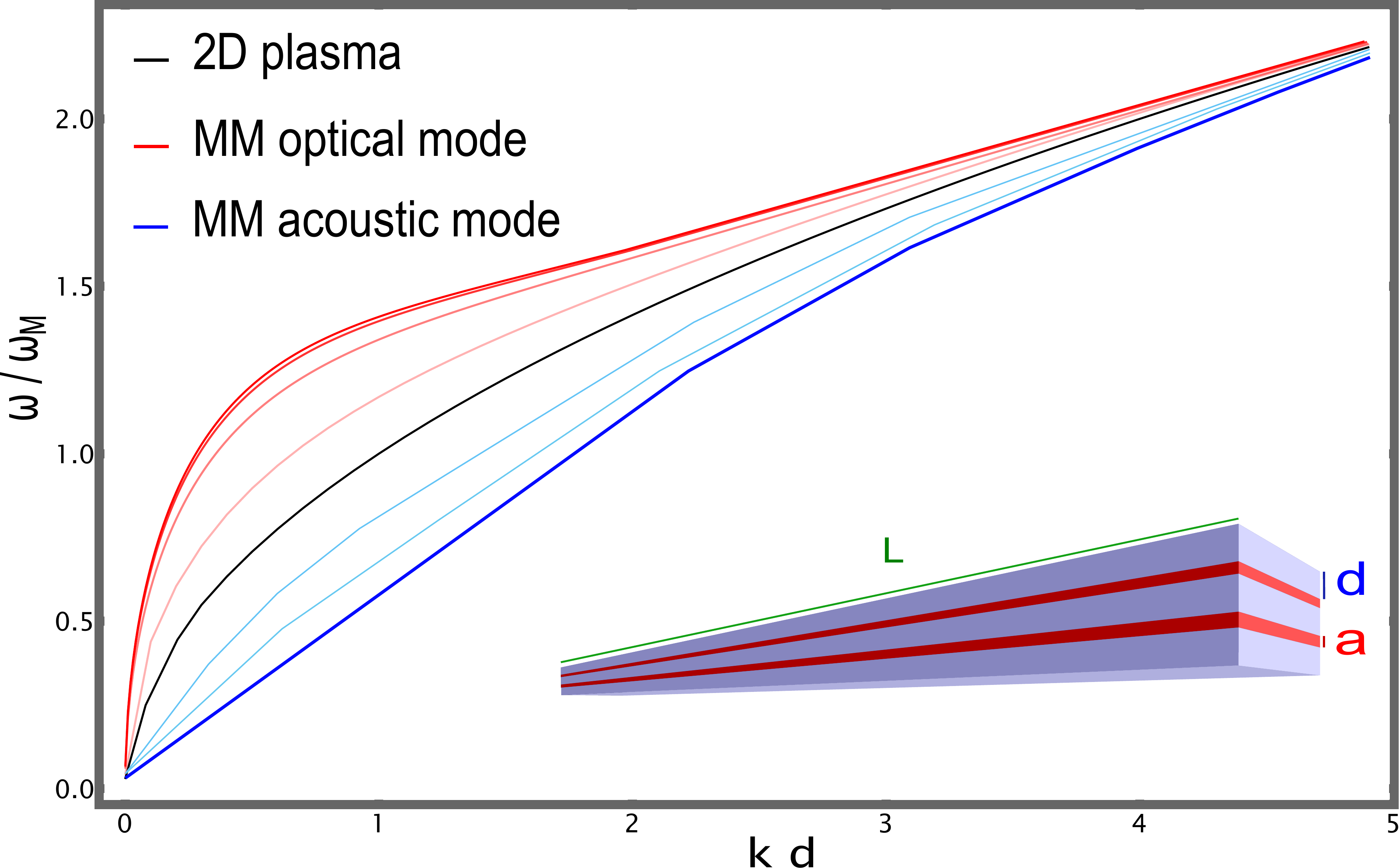}
\caption{The evolution of the mode from single bilayer graphene to metamaterial, the solid curves are obtained if the number of layers is larger than 20; in the inset a pictorial representation of the metamaterial.}
\label{fig:MMDR}
\end{figure}

\textit{A graphene metamaterial.} So far, we consider the interaction of a neutrino flux with a single bidimentional plasma, the instability occurs when the neutrinos travel through the material with an in-plane velocity; this event is high unlikely and therefore we need to enlarge the cross-section of the process. We thus stack several graphene layers spacing them by a thickness $d$ i.e., we created an effective tridimentional metamaterial. We aim to a simple model that describes the electrons behavior in the metamaterial as function of the number of layer and spacing distance. We consider a \textit{bilayer graphene} as basic element, which has an electron effective mass weakly dependent on the band structure \cite{zou:emehbg}, thus allowing us to write $M \simeq 0.02 \, \rm{}MeV$. Moreover we model the buffer layer as a perfect insulator with dielectric constant equals to one. 
The metamaterial dispersion relation is then calculated by considering the electrostatic interaction between the bidimentional electron distributions. Increasing the number of layer, the metamaterial dispersion relation converges showing an acoustic mode and an optical mode (see Fig. \ref{fig:MMDR}).
For higher wavevector, the modes behaves similarly to the bilayer graphene case; conversely, for low wavevectors, the metamaterial plasmon dispersion deviates from the bidimensional case. The acoustic mode is linear, while the optical mode has a typical square-root behavior with a larger group velocity than the bidimensional case. We focus on the optical mode in the low wavevector limit, therefore we can Taylor expand and write the metamaterial dispersion relation for the optical mode as
\begin{equation}
	\tilde{\omega}^2 - \frac{(36 - 35 \tilde{k})^2}{216} \tilde{k} =0,
	\label{eq:seriesMMDR}
\end{equation}
where $\tilde{k}=k d$, $\tilde{\omega}=\omega/\omega_M$, with $\omega_M^2=N_M e^2/2 M \epsilon_0$, and $N_M=N_e/d$ being the effective metamaterial density.

\begin{figure}[t!]
	\centering
	\includegraphics[width=0.5\textwidth]{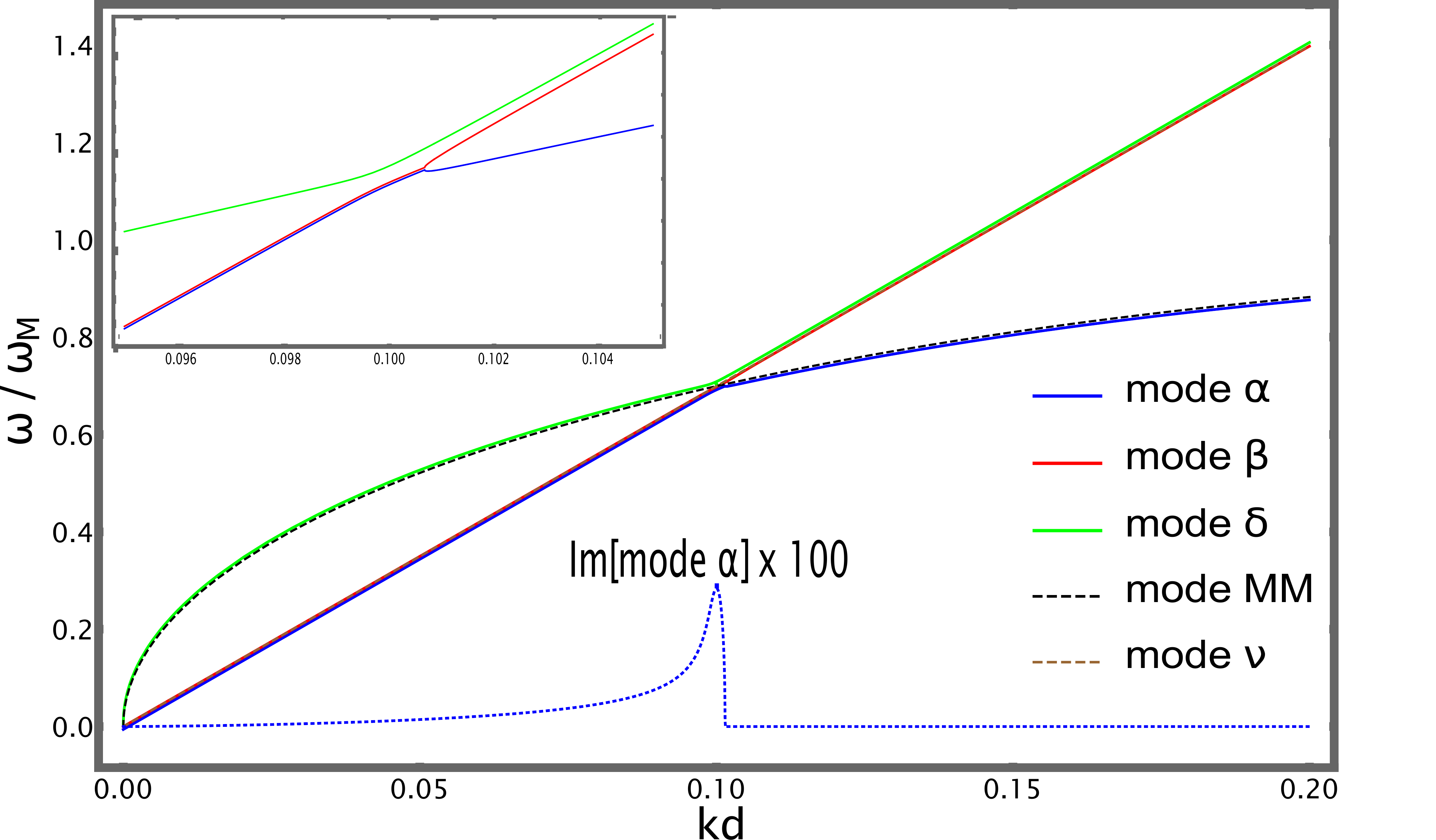}
	\caption{The joint dispersion relation of the neutrino-plasma interaction in a graphene metamaterial. In the inset, zoom around the crossing point, putting in evidence the bending of the bare modes. These modes are obtained considering $N_e=10^{15}\, \rm{m^{-2}}$, $N_\nu=10^4\, \rm{m^{-3}}$, $\beta_\nu= 0.1$ $m = 0,01\, \rm{eV}$ and imposing the crossing at $k_C=0.1$.}
	\label{fig:JDRmodes}
\end{figure}

\begin{figure}[t!]
	\centering
	\includegraphics[width=0.45\textwidth]{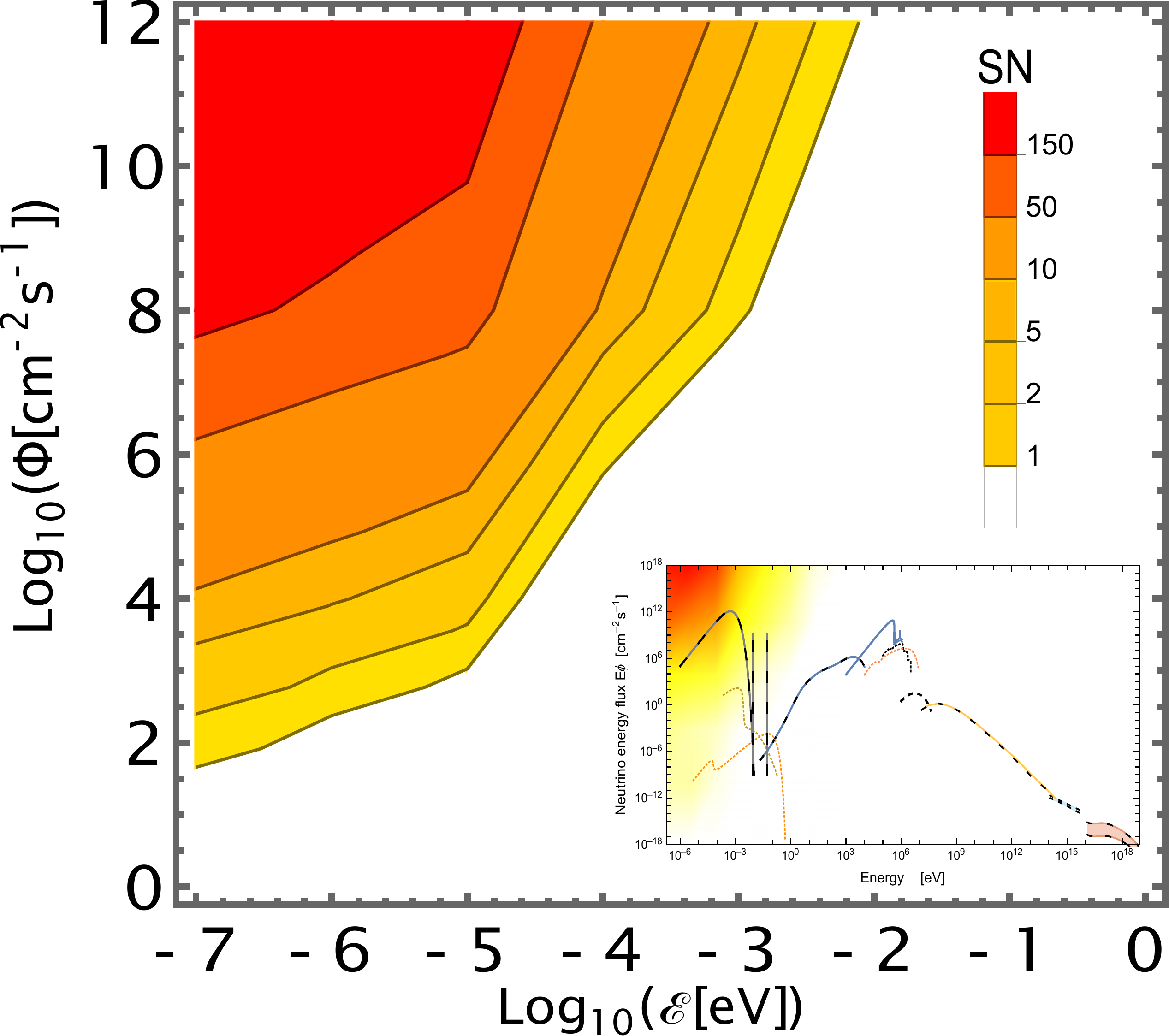}
	\caption{Evolution of the signal-to-noise ration for a graphene metamaterial of $L=1\, \rm{cm}$, a neutrino mass of $m=0.01\, \rm{eV}$ and for $N_e$ that increases from $10^{11}\, \rm{m^{-2}}$ to $10^{16}\, \rm{m^{-2}}$ as the energy increases.}
	\label{fig:SNfinal}
\end{figure}

\textit{Projection of the neutrino detector.} The interaction is now considered between the neutrino-flux and the electron plasma in the metamaterial. We  can thus introduce the coupling constant $\Gamma= \tilde{G}_{\rm F}^2 N_\nu N_M/m M c^4$, we further define the dimensionless velocities as 
\begin{equation}
	\tilde{V}_\nu= \frac{V_\nu}{d \, \omega_M} \;\; {\rm{and}} \;\; \tilde{c}_\nu= \frac{c}{d \, \omega_M}. 
	\label{eq:adimentionalVelocity}
\end{equation}
The interaction of the neutrino flux with the metamaterial plasma is described by \eqref{eq:matrixUngated} where the element $[1,1]$ is substituted by the Eq. \eqref{eq:seriesMMDR} and $N_e$ ($\tilde{g}_{\rm{F}}$) becomes $N_M$ ($\tilde{G}_{\rm{F}}$) therefore, imposing the determinant of such matrix to vanish, we get
\begin{equation}
	\left[\tilde{\omega}^2 - \tilde{k} \frac{(36 - 35 \tilde{k})^2}{216}\right]  (\tilde{\omega}-  \tilde{V}_\nu  \tilde{k} )^2 - \Gamma(\tilde{\omega}^2-  \tilde{c}^2  \tilde{k}^2)^2=0
	\label{eq:JDR}
\end{equation}
The solution to Eq. \eqref{eq:JDR} produces the modes shown in Fig. \ref{fig:JDRmodes}, where we highlight the imaginary part of mode $\alpha$ since it is responsible for the instability.
The growth rate, $\gamma = \rm{Im}({\rm{mode}}\, \alpha)$, increases when the coupling parameter $\Gamma$ increases, assuming $\tilde{V}_\nu$ does not change. We will focus on the plasmon that grows faster, it has wavevector $k_{\rm{C}}$ and frequency given by $\omega_R + i \gamma_{\rm{m}}$.
The complex frequency implies that the plasmon density not only oscillates but also grows exponentially as
\begin{equation}
	n_e(x,t)=n_e^0 \exp[i(k_C x - \omega_R t)] \exp(\gamma_{\rm{m}} t),\label{eq:plasmonExpression}
\end{equation}
where $n_e^0$ is the amplitude given by the thermal noise. We require that the plasmon amplitude grows of a factor hundred due to the instability, this implies 
\begin{equation}
	\rm{SN}_{\rm{dB}} = 10 \log_{10}\left[ \exp \left(2 \frac{\gamma_{m} L}{V_\nu}\right)\right] = 20,
\end{equation}
where we consider the neutrinos interact with the metamaterial for a time $t_D \simeq L/V_\nu$ i.e., a longer metamaterial produces then a larger plasmon. This request sets the length of the MM as function of the neutrino density and velocity. In case of neutrinos at $V_\nu= 0.1 \, c$ and $N_\nu= 10^4 \rm{m^{-3}}$, we need a metamaterial about $1 \, \rm{cm}$ long; reducing the density to $N_\nu= 10^{-2} \, \rm{m^{-3}}$ the required length increases to approximately $1 \, \rm{m}$. 
The neutrino flux can have different energy and the instability growth is significantly dependent on the neutrino velocity. Keeping the neutrino mass at ${m=0.01 \rm{eV}}$, we compute the signal to noise ratio in case of $1 \, \rm{cm}$ for a large range of energy and flux; the result is presented in Fig.\ref{fig:SNfinal}. 
It is straightforward to see that 
the neutrino-plasma instability is a technique suitable for the detection of ultra-low energetic neutrinos with large fluxes.

\textit{Conclusions.} We studied the interaction between neutrinos and electrons; from the quantum field theory interaction, we obtained, by the semiclassical approximation, the interaction lagrangian as function of density and velocity profiles.
We derived the force generated by a distribution on a single particle of the other species; the force shows a ponderomotive-like behavior and it is proportional to the Fermi constant. Using the kinetic description we coupled a neutrino   flux with a solid-state plasma. The plasma is made by the electrons in a graphene metamaterial, which is obtained by staking several bilayer graphene spaced by a buffer layer. The bare dispersion relation of such plasma is tunable by the inter-layer distance; in the limit of infinite number of layer, we obtained an optical mode. The metamaterial mode has a root square dependence and it crosses the linear dispersion relation of the neutrinos.
We worked within the first-order perturbation theory, obtaining the joint neutrino-plasma dispersion relation, which contains the hybridized modes of the neutrino and electron systems. We obtained a complex mode (instability), in which the imaginary part peaks at the wavevector at which the bare modes cross (resonance condition). Measuring the produced plasmon allows us to infer about the neutrinos properties, namely its energy, density and mass. This so happens since frequency of the excited plasmon depends on the neutrinos speed, while the amplitude and the spectral width are functions of the neutrinos density and mass. The neutrino-plasma instability is particularly efficient for ultra-low energetic neutrinos, therefore constituting an appealing way towards the detection of such elusive particles. The detection of such ultra-low energetic neutrinos may set novel boundaries on neutrino phenomenology and future particle-physics theories, with important impacts on eventual extensions of the Standard Model, dark matter searches and cosmology.
This work has to be understood as an important first step on the quest for slow neutrinos based on plasma physics phenomenology, therefore involving the collective response of the electrons. This is in strong contrast with the techniques employed in the existing schemes, such as IceCube or Super-Kamiokande, which are based on single-particle processes. Moreover, our findings challenge solid-state communities and the nano-technology of graphene devices, as our detection scheme will require a huge control of controllability of the next-generation graphene metamaterials. \par

\textit{Acknowledgments.} One of the authors (H.T.) acknowledges Funda\c{c}\~{a}o da Ci\^{e}ncia e Tecnologia (FCT-Portugal) through Contract No. CEECIND/00401/2018, and through the Project No. PTDC/FIS-OUT/3882/2020.
     
\bigskip

\bibliographystyle{apsrev4-1}
\bibliography{main_v2.bib}

\end{document}